\documentclass[12pt]{article}
\usepackage{epsfig}
\begin{document}
\begin{titlepage}
\pagestyle{empty} \baselineskip=28pt \rightline{\tt MZ-TH/05-02} 
\rightline{\tt AS-ITP-05-01}
\vskip 0.7in
\begin{center}
{\large {\bf Note on the Slope Parameter of the Baryonic \\
$\Lambda_b\to\Lambda_c$ Isgur-Wise Function}}
\end{center}
\begin{center}
\vskip 0.2in {\bf Ming-Qiu~Huang}$^1$, {\bf Hong-Ying~Jin}$^{2}$,
{\bf J.~G.~K\"orner}$^{3}$ and {\bf Chun~Liu}$^{4}$ \vskip 0.1in
{\it $^1${Physics Department, Nat'l University of Defense
Technology, Changsha 410073, China}\\
$^2${Institute of Modern Physics, Zhejiang University, 
Hangzhou 310027, China }\\
$^3${Institut f\"ur Physik, Johannes Gutenberg-Universit\"at,
D-55099 Mainz , Germany}\\
$^4${Institute of Theoretical Physics, Chinese Academy of
Sciences, \\P.O. Box 2735, Beijing 100080, China}}\\
\vskip 0.2in {\bf Abstract}
\end{center}
\baselineskip=18pt \noindent

Using the framework of the Heavy Quark Effective Theory we have 
re-analyzed the Isgur-Wise function describing semileptonic 
$\Lambda_b\to\Lambda_c$ decays in the QCD sum rule approach. The slope 
parameter of the Isgur-Wise function is found to be 
$\rho^2= 1.35\pm 0.12$, which is consistent with an experimental 
measurement and a lattice calculation.  To ${\cal O}(1/m_b,1/m_c)$ of 
the heavy quark expansion the integrated $\Lambda_b$ decay width is   
used to extract the CKM matrix element $V_{cb}$ for which we obtain a 
value of $|V_{cb}| = 0.041\pm 0.004$ in excellent agreement with the 
value of $|V_{cb}|$ determined from semileptonic $B\to D^*$ decays.  
\vfill \leftline{February 2005}
\end{titlepage}
\baselineskip=18pt
The study of $b\to c$ semileptonic weak decays has been the subject of 
considerable interest in recent years, as a source of information on 
$V_{cb}$ and as a laboratory for understanding the strong interaction 
effects and developing nonperturbative QCD methods. A considerable 
amount of work has been carried out in the meson sector, in which the 
Heavy Quark Effective Theory (HQET) \cite{hqet} was first developed. 
In the baryon sector a particular example is the semileptonic decay 
$\Lambda_b\to\Lambda_c\ell\bar\nu_\ell$.  In view of the fact that a 
recent experiment from DELPHI \cite{exp1} shows a discrepancy with the 
results of previous QCD sum rule calculations \cite{gy,dai1} an 
updated sum rule analysis for the baryonic $\Lambda_b\to\Lambda_c$ 
form factor is called for.  

In the heavy quark limit, the hadronic matrix element of the 
$\Lambda_b\to\Lambda_c$ transition can be simply expressed in terms of 
a single Isgur-Wise (IW) function defined as follows 
\cite{baryon-IW,hkkp,KKP},
\begin{equation}
\langle\Lambda_c(v')|\bar{c}\Gamma b|\Lambda_b(v)\rangle =
\xi (\omega)\bar{u}_{\Lambda_c}(v')\Gamma u_{\Lambda_b}(v)\,,
\end{equation}
where $\omega=v\cdot v^\prime$ is the velocity transfer variable and 
$\Gamma$ is an arbitary gamma matrix.  When the velocity of the heavy 
quark changes from $v$ to $v'$ due to the weak decay of the heavy 
quark, the light degrees of freedom undergo a corresponding transition 
due to their strong interactions with the heavy quark.  The 
Isgur--Wise (IW) function $\xi(\omega)$ is a measure of the transition 
amplitude of the light degrees of freedom.  The IW function 
$\xi(\omega)$ is normalized to 1 at the zero recoil point $\omega=1$.  
This value is reduced by a few percent after taking into account 
radiative QCD correction effects \cite{neubert-r,cm97} 
\footnote{At zero recoil the $\Lambda_b\to\Lambda_c$ weak transition 
matrix elements have no $1/m_Q$ corrections, and the $1/m_Q^2$ 
corrections are small.}  To obtain a theoretical description of the 
whole IW function one must use non-perturbative methods which, at the 
current stage, are beset with large uncertainties.  In the decay 
$\Lambda_b\rightarrow\Lambda_c\ell\bar\nu_\ell$, the physical region 
of $\omega$ lies in the range 1 to 1.43.  Usually the IW function is 
expanded up to the first order in $(\omega-1)$, 
\begin{equation}
\xi(\omega)=1-\rho^2(\omega-1)+{\mathcal O}((\omega-1)^2)
\end{equation}
The slope parameter ${\rho}^2$ of the IW function at zero recoil has 
to be calculated using nonperturbative methods.  

The QCD sum rule approach \cite{4a,hqetsum} has proven to be a 
reliable tool to deal with many problems in the realm of 
nonperturbative QCD.  It has been used successfully to calculate the 
properties of various hadrons.  For instance, besides the light mesons 
and baryons, heavy meson properties were systematically analyzed in 
the sum rule approach within the framework of HQET \cite{neubert-r}.  
In the heavy baryon sector the masses of heavy baryons and the IW 
functions describing their weak transitions were calculated in Refs. 
\cite{gy,dai1,sky,5a0} and \cite{grozin,dai2,wh,konig97,ikkl97}, 
respectively.  In Ref. \cite{5a}, the calculation for the heavy 
baryons began with the full theory and the results of the calculation 
were expanded in terms of inverse powers of the heavy quark masses. In 
the HQET sum rule approach the baryonic $\Lambda_b \to\Lambda_c $ IW 
function was calculated in \cite{gy,dai1}, and the slope parameter 
$\rho^2$ was fitted to lie in the range $0.5-0.8$.  However, such low 
slope values  would predict exclusive $\Lambda_b\to\Lambda_c$ 
semileptonic decay rates dangerously close to the inclusive 
semileptonic rate \cite{li,melic}. A first measurement of the IW 
function of semileptonic $\Lambda_b \to \Lambda_c$ transitions has 
recently been reported by the DELPHI Collaboration \cite{exp1}. The 
errors on this measurement are quite large.  Using an exponential 
parametrisation, they quote a value of 
$\rho^2=1.59\pm 1.10 (\mbox{stat})$.  When the observed event rates 
were included in the fit they obtained 
$\rho^2=2.03\pm 0.46(\mbox{stat})^{+0.72}_{-1.00}(\mbox{syst})$ 
\cite{exp1}.  Within the large error bars the experimental slope value 
is compatible with the HQET sum rule results of \cite{gy,dai1} 
although the experimental central values of \cite{exp1} are 
considerably higher than the theoretical sum rule results.

Theoretically, the above-mentioned HQET sum rule results for the slope 
parameter $\rho^2 = 0.5 - 0.8$ appear to be rather small.  Because the 
number of light quark transitions is larger in the heavy baryon case 
than in the heavy meson case one expects that the slope of the 
baryonic IW function is larger than that of the mesonic IW function.  
In fact, in the large $N_c$ limit, $\rho^2$ will be infinitely large 
\cite{7}. In the spectator quark model approach \cite{hkkp,KKP,9} one 
finds $\rho^2 =2\rho^2_{\mbox{meson}}-1/2$ which turns into an upper 
bound $\rho^2 \le 2\rho^2_{\mbox{meson}}-1/2$ when the interaction 
between the light quarks is turned on \cite{iklr99} \footnote{The 
large $N_c$ result in the spectator quark model approach can be worked 
out to be $\rho^2 =(N_c-1) (\rho^2_{\mbox{meson}}-1/4)$.  Since 
$\rho^2_{\mbox{meson}} \ge 1/2$ according to the sum rule of Bjorken, 
one then recovers the infinite slope result of \cite{7} in the large 
$N_c$ limit.}.

Concerning the slope parameter of the mesonic IW function, one finds 
theoretical values of about $1$ from sum rule calculations 
\cite{neubert-r}. Experimental numbers for the mesonic slope parameter 
also scatter around 1 \cite{vcb,belle,babar}. Using the spectator 
quark model estimate one thus expects baryonic values of the slope 
parameter $\rho^2$ in the vicinity of 1.5 or slightly below that 
number.  The Skyrme model predicts $\rho^2\simeq 1.3$ \cite{8}. In the 
infinite momentum frame model one has $\rho^2 = 1.44$ \cite{konig97} 
and in the relativistic three quark model one finds $\rho^2 = 1.35$ 
\cite{ikkl97}. For the baryonic sum rule results, radiative 
corrections to $\rho^2$ and $1/m_Q$ corrections to the form factors 
are not expected to be large enough to solve the discrepancy between 
the large experimental central value for $\rho$ in \cite{exp1} and the 
small QCD sum rule results \cite{gy,dai1}.  We therefore concentrate 
only on the leading order results in our analysis.

The purpose of this work is to present a new QCD sum rule analysis for 
the leading-order Isgur-Wise function describing the 
$\Lambda_b\to\Lambda_c$ transition. In particular, we concentrate on 
the sum rule prediction directly for the slope parameter $\rho^2$.  To 
start with, we first review the sum rule analysis of the two-point 
Green's function involving two heavy baryon currents relevant for the 
determination of the heavy baryon decay constant and its mass.  A 
possible choice of the heavy baryon current with the correct quantum 
numbers of the heavy baryon $\Lambda_Q$ is given by
\begin{equation}
\label{current}
\tilde{j}^v=q^{\rm T}C\tau\gamma_5 q h_v \; ,
\end{equation} 
\noindent where $C$ is the charge conjugation matrix, $\tau$ is an 
antisymmetric flavor matrix, and $h_v$ and $q$ are the heavy and light 
quark fields.  Note that there is an alternative choice for the heavy 
baryon current $\tilde{j}^v$ which is obtained by the replacement 
$\gamma_5 \to \gamma_5 \rlap/ v $ in Eq.(\ref{current}).  The two 
baryon currents give the same diagonal sum rule for the IW function.  

>From the correlator
\begin{equation}
\label{01} 
\Gamma(\epsilon)=i\int\;d^4x\; e^{ikx}\langle
0|T\{\tilde{j}^v(x)\overline{\tilde{j}^v}(0)\}|0\rangle\,
\end{equation}
one can obtain the two-point sum rule \cite{gy,dai1,sky,5a0}:
\begin{equation}
\label{02}
2 f^2 e^{-\bar{\Lambda}/T} = \frac{1}{10\pi^4}\int_0^s d\nu \nu^5
e^{-\nu/T} + \frac{\langle\bar{q}q\rangle^2}{3}e^{-\frac{m_0^2}{8T^2}}
+ \frac{\langle\alpha_s GG\rangle}{16\pi^3}T^2\;,
\end{equation}
where the baryonic decay constant is defined as 
$\langle0|\tilde{j}^v|\Lambda_Q\rangle\equiv f u$.  
$\bar{\Lambda}=m_{\Lambda_Q}-m_Q$ is the binding energy of the 
$\Lambda_Q$--baryon in HQET.  $T$ is the Borel parameter.  According 
to the duality assumption the higher resonance and continuum 
contributions to the Green's function are approximated by the 
perturbative contribution above a given threshold $s$.  Note that the 
factor $2$ on the l.h.s. of Eq. (\ref{02}) is missing in the 
calculation of Ref.\cite{grozin}. The missing factor $2$ is, however, 
of no relevance for the calculation of $\bar{\Lambda}$ and the IW 
function.

To obtain the IW function, one considers the three-point correlator
\begin{equation}
\label{three-point}
\label{cor} \tilde{\Xi}(\epsilon,\epsilon',\omega)=
i^2\int d^4x_1 d^4x_2\;e^{i(k'x_1-kx_2)} 
\langle 0|T\tilde{j}^{v'}(x_1)\bar{h}_{v'}(0)\Gamma
h_v(0) \overline{\tilde{j}^v}(x_2)|0\rangle\,,
\end{equation}
where $\epsilon=v \cdot k$ and $\epsilon'=v' \cdot k'$.
In the hadronic language the three-point function 
Eq. (\ref{three-point}) can be expressed in terms of insertions of 
hadronic states. The lowest contribution involves the IW function,
\begin{equation}
\label{cor1} 
\tilde{\Xi}(\epsilon,\epsilon',\omega)=
\frac{f^2\xi(\omega)}
{(\bar{\Lambda}-\epsilon)(\bar{\Lambda}-\epsilon')}
\frac{1+\not\! v'}{2}\Gamma\frac{1+\not\! v}{2}+ {\rm resonances}\,.
\end{equation}
On the other hand, the three-point correlator $\tilde{\Xi}$ can be 
calculated by using the Operator Product Expansion.  The perturbative 
contribution can be written in terms of the double dispersion 
relation,
\begin{equation}
\label{cor2}
\tilde{\Xi}^{\rm pert}(\epsilon,\epsilon',\omega)=
\int_0^\infty d\tilde{\epsilon}\int_0^\infty d\tilde{\epsilon'}
\frac{\varrho(\tilde{\epsilon}, \tilde{\epsilon'}, \omega)}
{(\tilde{\epsilon}-\epsilon)(\tilde{\epsilon'}-\epsilon')}\,,
\end{equation}
with the following spectral density,
\begin{equation}
\label{cor3} \varrho(\epsilon,\epsilon',\omega)=
\frac{3}{2^6\pi^4}\frac{1}{(\omega^2-1)^{5/2}}(\epsilon^2+\epsilon'^2
-2\omega \epsilon\epsilon')^2{\rm tr}(\tau\tau^\dagger)
\frac{1+\not\! v'}{2}\Gamma\frac{1+\not\! v}{2}\Theta(\epsilon)
\Theta(\epsilon') \Theta(2\omega\epsilon\epsilon'-\epsilon^2-
\epsilon'^2)\,.
\end{equation}
The condensate contributions will be included as a series of vacuum 
expectation values of operators ordered by their dimension.

A systematic uncertainty of QCD sum rule calculations lies in the
treatment of higher state contributions to the hadronic side of
$\tilde{\Xi}$.  Generally the quark-hadron duality assumption is
adopted, which simulates the higher state contribution by the
perturbative part above some threshold energy.  For three-point 
Green's functions, this assumption is more ambiguous than that for the 
two-point case because there are two energy variables 
$\tilde{\epsilon}$ and $\tilde{\epsilon}'$. It was argued by Blok and 
Shifman in \cite{bshifman}, that the perturbative and the hadronic 
spectral densities cannot be locally dual to each other. The necessary 
way to restore duality is to integrate the spectral densities over the
``off-diagonal'' variable
$\tilde{\epsilon}_-=(\tilde{\epsilon}-\tilde{\epsilon}')/2$, keeping
the ``diagonal'' variable
$\tilde{\epsilon}_+=(\tilde{\epsilon}+\tilde{\epsilon}')/2$ fixed. It
is w.r.t. $\tilde{\epsilon}_+$ that the quark-hadron duality is 
assumed for the integrated spectral densities. With this procedure the 
sum rule for the IW function yields \cite{dai2},
\begin{equation}
\label{cor4}
2 f^2 \xi e^{-\bar{\Lambda}/T} = \frac{4}{5\pi^4}
\frac{1}{(\omega+1)^3}\int_0^{\tilde{s}}d\tilde{\epsilon_+}
\tilde{\epsilon_+}^5e^{-\tilde{\epsilon_+}/T}
+\frac{\langle\bar{q}q\rangle^2}{3}e^{-\frac{m_0^2}{16T^2}(\omega+1)}
+ \frac{\langle\alpha_s GG\rangle}{12\pi^3}T^2
\frac{2\omega+1}{(\omega+1)^2}\,,
\end{equation}
where $\tilde{s}$ is the threshold energy for the sum rule.  It should
be noted that the Borel parameter has been chosen such that 
$\xi(\omega=1)=1$. The $\omega$ dependence in the gluon condensate 
term and the coefficient of the exponential in the quark condensate 
term are different from Ref. \cite{grozin}, but are consistent with 
Refs. \cite{wh,5a}. From Eq. (\ref{cor4}) one obtains the sum rule for 
the slope parameter of the IW function, i.e. 
$\displaystyle \rho^2\equiv-\frac{d\xi}{d\omega}|_{\omega=1}$.  It is 
given by
\begin{equation}
\label{cor5} 
2 f^2 \rho^2 e^{-\bar{\Lambda}/T} = \frac{3}{20\pi^4}
\int_0^{\tilde{s}}d\tilde{\epsilon_+}\tilde{\epsilon_+}^5
e^{-\tilde{\epsilon_+}/T} +
\frac{m_0^2\langle\bar{q}q\rangle^2}{48T^2}e^{-\frac{m_0^2}{8T^2}} +
\frac{\langle\alpha_s GG\rangle}{48\pi^3}T^2 \,.  
\end{equation}
The  baryonic decay constant $f$ can be obtained by the sum rule 
Eq. (\ref{02}).  Note that we have not included the perturbative 
${\cal O}(\alpha_s)$ corrections in the sum rule which is expected to 
be largely cancelled in the ratio of Eqs. (\ref{cor4}) and (\ref{02}), 
or Eqs. (\ref{cor5}) and (\ref{02}).  This is what happened in the 
case of heavy meson form factors \cite{neubert-r}.

For the numerical analysis of the sum rule (\ref{cor5}) we use the 
two-point sum rule (\ref{02}) to eliminate the explicit dependence of 
Eq. (\ref{cor5}) on $f$ and $\bar\Lambda$. This procedure reduces the 
uncertainties in the calculation. For the condensate contributions we 
take the standard values
\begin{eqnarray}
\label{cond}
\langle\bar q q\rangle=-(0.24\pm 0.01)^3~\mbox{GeV}^3\,,\hspace{3mm}
\langle{\alpha_s\over\pi} GG\rangle=(0.012\pm0.004)~\mbox{GeV}^4 \,
\end{eqnarray}
and $m_0^2=0.8$ GeV$^2$. The threshold energies are taken to be equal, i.e.
$s=\tilde{s}=s_c$. Imposing the usual criterion on the ratio of 
contributions of the higher-order power corrections and that of the 
continuum and using the central values of the condensates given in 
(\ref{cond}) there is an acceptable window of stability in the range 
$T=0.4-0.7$ GeV in which the calculated results do not change 
appreciably if the threshold parameter $s_c$ lies in the range 
$1.7<s_c<2.1$ GeV. In Fig. 1, the sum rule for the slope parameter of 
the IW function $\rho^2$ is plotted as a function of the Borel 
parameter $T$ for various choices of the continuum threshold in the 
range $1.7<s_c<2.1$ GeV.  One can see that the variation is very 
moderate for the Borel parameter in the range $0.4<T<0.7$ GeV.  Our 
prediction for the slope parameter $\rho^2$ is given by
\begin{eqnarray}
\label{value}
\rho^2=1.35\pm0.12 \,.
\end{eqnarray}
\begin{figure}[t]
\centerline{\epsfysize=6.5truecm \epsfbox{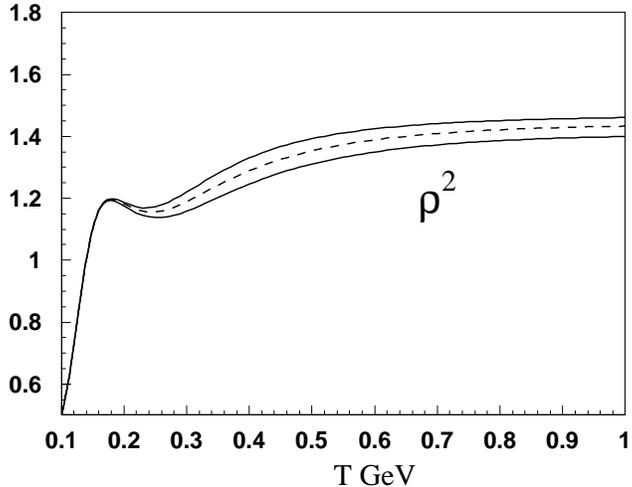}} 
\caption{The slope parameter of the IW function $\rho^2$ versus the 
Borel variable $T$.  The curves refer to choices of the threshold 
parameters: $s_c=2.1$, $s_c=1.9$, $s_c=1.7$, from top to bottom.} 
\label{fig}
\end{figure}
The errors reflect the uncertainty due to the sum rule window.  The
value of the slope parameter is in agreement with the recent 
experimental result \cite{exp1} and the value obtained in a lattice 
determination \cite{lattice}.

It is meaningful to ask why the present sum rule result is so 
different from those obtained previously.  First, this maybe due to 
the systematic uncertainty of QCD sum rules.  The systematic error 
resulting from the use of quark-hadron duality above $s_c$ is 
difficult to estimate. Conservatively speaking, there is a $10\%-30\%$ 
systematic error. Second, the previous linear fit to the IW function 
\cite{dai2,grozin} may be a rather poor fit in the $\omega$ range: 
$1-1.43$.  In the following decay rate calculation, we assume an 
exponential form to parametrize the baryonic IW function, as the 
experiment did \cite{exp1}:
\begin{equation}
\label{expo}
\label{slo} \xi(\omega)=\xi(1)\;\mbox{exp}[-\rho^2(\omega-1)] \;.
\end{equation}

Once we have computed the IW function we are now in the position to 
calculate the rate for the semileptonic 
$\Lambda_b\to\Lambda_c\ell\bar\nu_\ell$ transitions.  Neglecting the 
lepton mass the differential decay rate can be written as 
(see e.g. \cite{dai1,KKP})
\begin{eqnarray}
\frac{d\Gamma}{d\omega}&=&\frac{G_{F}^{2}|V_{cb}|^{2}m_{\Lambda_b}^{5}
r^{3}}{24\pi^{3}}\sqrt{\omega^2-1}\bigg\{(\omega-1)\bigg[2\kappa 
F_{1}^{2}+[(1+r)F_1+(\omega+1)(rF_2+F_3)]^2\bigg] \nonumber \\&& 
+(\omega+1)\bigg[2\kappa G_{1}^{2} + [(1-r)G_1-(\omega-1)(rG_2+G_3)]^2
\bigg]\bigg\}\;,
\end{eqnarray}
where $r=m_{\Lambda_c}/m_{\Lambda_b}$ and $\kappa=1+r^2-2r\omega$. The form
factors $F_i$ and $G_i$ can be expressed by a set of Isgur-Wise 
functions at each order in $1/m_Q$ in HQET. Taking into account the 
$1/m_Q$ corrections, they are (see e.g. \cite{dai1,KKP})
\begin{eqnarray}
&&F_1=[1+(\varepsilon_c+\varepsilon_b)\bar\Lambda]\;\xi(\omega)\;,
\hspace{14.2mm}
F_2=G_2=-\frac{2\varepsilon_c\bar\Lambda}{\omega+1}\;\xi(\omega)\;,
\nonumber\\
&&G_1=[1+(\varepsilon_c+\varepsilon_b)\bar\Lambda
\frac{\omega-1}{\omega+1}]\;\xi(\omega)\;,\hspace{3mm}
F_3=-G_3=-\frac{2\varepsilon_b\bar\Lambda}{\omega+1}\;\xi(\omega)\;,
\end{eqnarray}
where $\varepsilon_Q=1/(2m_Q)$. Notice that the subleading Isgur-Wise 
function associated with the insertion of the $\Lambda_{\rm QCD}/m_c$ 
kinetic operator of the HQET Lagrangian has been neglected since it is 
negligibly small \cite{dai2,konig97}.  In the numerical calculation we 
take the heavy quark masses to be $m_b=4.8$ GeV, $m_c=1.4$ GeV and 
$\bar\Lambda=0.79$ GeV \cite{dai1}.  With the masses of $\Lambda_b$ 
and $\Lambda_c$ given by the Particle Data Group (PDG) \cite{PDG} the 
upper limit of $\omega$ is $\omega_{max}=(1+r^2)/2r=1.433$.  The decay 
rate can then be calculated to be
\begin{eqnarray}
\Gamma(\Lambda_b\to\Lambda_c\ell\,\bar\nu_\ell)&=&2.12\times
10^{-11}\;|\mbox{V}_{cb}|^2\; \mbox{GeV}\;.
\end{eqnarray}
The contribution of the $1/m_Q$ corrections amount to about $10\%$.

Let us compare our theoretical prediction to experimental
data. Recently, the DELPHI Collaboration measured the
$\Lambda_b\to\Lambda_c\ell\,\bar\nu_\ell$ branching ratio \cite{exp1} 
as 
${\cal B}(\Lambda_b\to\Lambda_c\ell\,\bar\nu_\ell)=(5.0^{+1.1}_{-0.8}
(\mbox{stat})^{+1.6}_{-1.2}(\mbox{sys}))\%$.  On the other hand, the 
branching ratio given by the PDG is 
${\cal B}(\Lambda_b\to\Lambda_c\ell\,\bar\nu_\ell+\mbox{anything})=
(9.2\pm2.1)\%$ \cite{PDG}.  An error weighted average value was given 
by Albertus {\it et al.} \cite{AHL}:
\begin{eqnarray}
\label{ewav}
{\cal B}(\Lambda_b\to\Lambda_c\ell\,\bar\nu_\ell)=(6.8\pm 1.3)\%.
\end{eqnarray}
The total $\Lambda_b$ decay width is determined by the inverse 
lifetime $\tau_{\Lambda_b}^{-1}$ where we take 
$\tau_{\Lambda_b}=1.229\pm 0.80$ ps \cite{PDG}.  The value of the CKM 
matrix element $|\mbox{V}_{cb}|$ can then be extracted to be 
\begin{eqnarray}
|\mbox{V}_{cb}|=0.041\pm 0.004\;,
\end{eqnarray}
where the error is due to experimental uncertainties. The value of 
$|\mbox{V}_{cb}|$ is in excellent agreement with the recent 
experimental determination by the DELPHI Collaboration 
$|V_{cb}|=0.0414\pm 0.0012(\mbox{stat})\pm0.0021(\mbox{syst})\pm
0.0018(\mbox{theory})$, obtained from semileptonic 
${B} \to D^*\ell{\bar\nu}_\ell$ decays \cite{vcb}.  It is also in good 
agreement with the value of $|\mbox{V}_{cb}|$ obtained from a 
nonrelativistic quark model calculation of the $\Lambda_b\to\Lambda_c$ 
transition \cite{AHL}.

In conclusion, we have presented a HQET sum rule analysis for the 
slope parameter $\rho^2$ of the baryonic IW function. We obtained  
$\rho^2= 1.35\pm 0.12$ which is in good agreement with the recent 
DELPHI measurement and the results of a lattice calculation. When 
combined with the error weighted average value for the branching ratio 
of $\Lambda_b\to\Lambda_c\ell\,\bar\nu_\ell$ quoted in 
Eq.(\ref{ewav}), our integrated decay width including $1/m_Q$ 
corrections leads to a value for $|V_{cb}|$ in excellent agreement 
with a recent determination of $|\mbox{V}_{cb}|$ from $B\to D^*$ 
decays.  One should bear in mind that $1/m_Q^2$ corrections can 
decrease the decay rate by a few percent as estimated in \cite{melic}.  
They should be studied within HQET sum rules in future works.  

\vskip 0.8in

{\bf Acknowledgment} We are grateful to Andrey Grozin  for very
helpful discussions.  Work on this project was begun while H.Y.J. and 
C.L. were fellows of the Alexander von Humboldt Foundation at the 
University of Mainz. It was also supported in part by the National 
Natural Science Foundation of China under Contract No. 10075068, 
10275091 and 10375057, and BEPC Opening Projects.

\end{document}